
plainTeX
\magnification=\magstep1
\font\bigbf=cmbx10 scaled\magstep1

\def\picture #1 by #2 (#3){
\vtop{\vskip #2
\special{picture #3}
\hrule width #1 height 0pt depth 0pt
\vfil}}
\def\scaledpicture #1 by #2 (#3 scaled #4){
\dimen0=#1 \dimen1=#2
\divide\dimen0 by 1000 \multiply\dimen0 by #4
\divide\dimen1 by 1000 \multiply\dimen1 by #4
\picture \dimen0 by \dimen1 (#3 scaled #4)}
\hoffset=.4truecm
\voffset=.5truecm
\hsize=15truecm
\vsize=23truecm
\baselineskip=12pt
\hfuzz=18pt
\newcount\eq
\def\adv{\global\advance\eq by1}
\newcount\eqncount\newcount\sectcount\eqncount=0\sectcount=0
\def\equn{\global\advance\eqncount
by1\eqno{(\the\eqncount)} }
\def\put#1{\global\edef#1{(\the\eqncount)}           }
\def\Bray{1}
\def\Mez{2}
\def\Fu{3}
\def\DeDominicisa{4}
\def\DeDominicisb{5}
\def\Garey{6}
\def\Par{7}
\def\Zinn{10}
\def\Kaz{9}
\def\Mott{11}
\def\Derrida{12}
\def\David{8}
\def\John{13}
{\nopagenumbers
\vbox{\vskip 2truecm
\centerline{\bigbf QUENCHED RANDOM GRAPHS}}
\vfil
\centerline{C. Bachas, C. de Calan and P.M.S.
Petropoulos}
\vskip 0.6cm
\centerline{\it Centre de Physique Th\'eorique, Ecole
Polytechnique\footnote{$^{\dag}$}{\sevenrm Laboratoire
Propre du Centre National de la Recherche Scientifique UPR
A.0014}}
\centerline{\it 91128 Palaiseau Cedex, France}
\vfil
\centerline{\bf Abstract}
\vskip 0.6cm
Spin models on quenched random graphs are related to many important
optimization problems. We give a new derivation of their mean-field
equations that elucidates the role of the natural order parameter
in these models.
\vfil
\par\noindent \line{CPTH-A264.1093
\hfil April 1994}
\eject}
\pageno=2
Spin models on quenched random graphs
have been studied extensively in recent
years [{\Bray}-\DeDominicisb] for a couple of reasons.
First, a large class of hard (and interesting in practice) optimization
problems such as graph partitioning and graph colouring [\Garey]
can be formulated as a search for the ground state
of such models. Their zero-temperature limit could thus yield valuable
information on average properties of the optimal solutions.
Second, for finite connectivity such models are closer to
realistic systems than their infinite-range counterparts, yet
mean-field theory is expected to stay exact.
They thus provide a simpler setting in which to try to test
whether the ultrametric structure and other
properties of Parisi's solution of the   spin-glass phase [\Par]
survive for finite-range interactions.
\vskip 0.3cm
In this letter we would like to give a new  derivation of the
mean-field equations [\DeDominicisa, \DeDominicisb]
for such models. It is based on some simple arguments, well-known
from the study of matrix models of 2d gravity [\David, \Kaz] and of the
large-order behaviour of perturbative series [\Zinn], and  adapted here
in the context of disordered systems.
Besides being simple and exact this novel derivation
elucidates the role of the natural
order parameter in these models
[\DeDominicisa, \DeDominicisb]. It can be furthermore adapted
readily to a variety of different situations.
We will not address here the hard problem of solving these equations
in the spin-glass phase. We will however
comment briefly on the phase diagram in the case   of
pure ferromagnetic or antiferromagnetic
couplings, as well as on the interpretation of the underlying
graphs as infinite-genus triangulations.
\vskip 0.3cm
Consider first the ensemble of all trivalent ($\phi^3$) graphs
made out of  $2n$ vertices.
If one ignores accidental-symmetry factors,  the number of such
graphs is  given by the integral expression
$$
{\cal N}_n = {1\over 2\pi i} \oint {d\lambda \over \lambda^{2n+1}}
\! \int_{-\infty}^{+\infty}
{d\phi\over \sqrt{2\pi}}\, e^{-{1\over 2} \phi^2 +
{\lambda\over 6}\phi^3}
                                          \ . \eqno(1)
$$
Indeed the $\phi$-integral can be expressed as a sum over all
topologically-distinct $\phi^3$ graphs weighted with
$\lambda^{\# {\rm vertices}}$,
times an inverse symmetry factor. The contour $\lambda$-integral then
picks out only the contribution of graphs with precisely $2n$ vertices.
In the large-$n$ limit we can evaluate this integral at the
dominant non-trivial saddle points for both variables
$\phi$ and $\lambda$.
After a   rescaling of variables $(\phi\to \phi/\lambda$)
and some straightforward Gaussian integrations  the result
of the  calculation reads:
$$
{\cal N}_n =
\left({n\over e}\right)^n
{\hat S}^{-n}
\left(-2\pi n  \det {\hat S}''\right)^{-1/2}
\big(1+o(1/n)\big) \ .                           \eqno(2)
$$
Here $S= {\phi^2\over 2}  - {\phi^3\over 6}$ is the
rescaled ``action" of the theory, ${\hat S}$ its value at the
dominant non-trivial saddle point ${\hat\phi}$
which solves the
``field equation"
$$
{\partial S \over \partial \phi}= 0 \ ,           \eqno(3)
$$
and $\det{\hat S}''$ is simply the second
derivative of $S$ at the saddle point.
Eq. (2) is a standard result for the large-order behaviour of
perturbative expansions [\Zinn],
and will stay valid in the more complicated cases studied
below. In the case at hand, using
${\hat\phi}=2$,
${\hat S} = 2/3$ and $-{\hat S}'' = 1$ one
recovers the correct counting of large undecorated   $\phi^3$
graphs, whose precise number is
${\cal N}_n = \left({1\over 6}\right)^{2n} {(6n-1)!! \over (2n)!}$.
\vskip 0.3cm
Let us consider now   an Ising model
with spins, $\sigma_{i}=\pm1$, lying on the $2n$ vertices of
a $\phi^3$ graph  ${\cal G}_n$. The
partition function is
$$
Z_{{\cal G}_n}(J,h) =\sum_{\sigma _{1},\ldots ,\sigma _{2n}}
\exp
\left( J\sum_{\langle ij \rangle} \sigma_i \sigma_j
+ h\sum_i \sigma_i \right) \ ,                    \eqno(4)
$$
where   the sums in
the Boltzmann weight run over all edges and vertices respectively of
the  graph ${\cal G}_n$, $J$ is the spin-spin coupling
and $h$   a magnetic field. The
average of the partition function over all graphs can be
expressed as an integral
[\Kaz]
over a ``field" defined on the discrete space $\{+,-\}$:
$$
{\overline {Z_{{\cal G}_n}(J,h)}}  \times {\cal N}_n   =
 {1\over 2\pi i} \oint {d\lambda\over \lambda^{2n+1}}\!
\int
{d\phi_+ \,d\phi_-\over 2\pi \sqrt{\det \Delta }}\,
\exp (-S)  \ ,                                       \eqno(5)
$$
where
$$
S = {1\over 2}
\sum_{\sigma,\tau} \phi_\sigma^{\phantom 3}
( \Delta^{-1})_{\sigma \tau}^{\phantom 3} \,
\phi_{\tau}^{\phantom 3}
-  {\lambda\over 6}
\left(e^{h}_{\phantom +} \phi_+^3+e^{-h}_{\phantom +}\phi_-^3\right)
                                             \ ,     \eqno(6)
$$
and the $2\times 2$ ``propagator" matrix  has entries
$$
\Delta_{\sigma \tau} = e^{J \sigma \tau}
                                                 \ . \eqno(7)
$$
Indeed, the  weak-$\lambda$ expansion of the
$\phi_\sigma$
integral(s)  is given as before by the sum over $\phi^3$
Feynman diagrams, while the $\lambda$-integration forces
the number of vertices to be $2n$.
For any given diagram the
vertices   are however now labelled by a ``position
in real space" $\sigma_i =\pm\, $. Furthermore there is a weight
$e^{h\sigma_i}$ for each vertex
and a propagator $e^{J\sigma_i \sigma_j}$ for each edge. Summing over all
``positions" of vertices thus yields the partition function of the Ising
model
on the corresponding graph. This justifies eq. (5). Note that
the $\phi_\sigma$ integral is strictly-speaking only defined through
its asymptotic expansion.
\vskip 0.3cm
In the thermodynamic limit of large graphs ($n \to \infty$) we can again
calculate the above integral   by the saddle-point technique.
We limit ourselves  for simplicity to the case of vanishing
magnetic  field. The   action, eq. (6),  has three non-zero
saddle points, which after  the usual rescaling read:
$$
{\hat\phi}_+ ={\hat\phi}_- = {2\sqrt{g}\over g+1}\  ,
                                              \eqno(8)
$$
and
$$
{\hat\phi}_{\pm} = {\sqrt{g}\over (g-1)}
\left( 1\pm \sqrt{g-3\over g+1} \right)  \ , \ \  { \rm or}  \ \
{\hat \phi}_+ \leftrightarrow {\hat \phi}_- \  .            \eqno(9)
$$
Here $g\equiv e^{2 J}$, so that
$g\in [\,1,\infty)$
corresponds to ferromagnetic couplings $J>0$, while $g\in
[\, 0,1]$ to antiferromagnetic couplings $J<0$.
The (degenerate) saddle points, eq. (9),   dominate in
the low-temperature ferromagnetic region $g>3$, but
become complex below $g=3$, where the
saddle point (8) takes over. This latter can be
continued analytically all the way down to $g=0$,
i.e. to the zero-temperature antiferromagnet.
The
transition at $g=3$ corresponds in fact to the onset of
ferromagnetic order. This can be seen from the expression
for the average (annealed)  magnetization:
$$
{\cal M}_{\rm ann.} \equiv {1\over 2n}
{\partial\over\partial h}\log \overline {Z_{{\cal G}_n}}
\bigg\vert_{h=0}=
{{\hat\phi}_+^3 - {\hat\phi}_-^3\over
{\hat\phi}_+^3 + {\hat\phi}_-^3}
= \cases{\pm
{g\over g-2}\sqrt{g-3\over g+1},&if $g> 3$;\cr
 0,&if $g< 3$\cr }
                                     \eqno(10)
$$
which follows  by straightforward manipulations.
For completeness we give
also the result for the average partition function,
valid up to terms of order $o(1/ n)$:
$$
\eqalignno{
\log \overline {Z_{{\cal G}_n}(g)}\big\vert_{h=0}
&= - n \log {3\over 2}{\hat S}
-{1\over 2} \log  \left( -
\det \big(\Delta {\hat S}''\big)\right)            &(11)\cr
=&  \cases{
- n \log
{2g\sqrt{g}\over (g+1)^3}  -{1\over 2}
\log{3-g\over g+1 }                  ,&if $g<3$;\cr
-n \log
{3\sqrt{3}\over 32} + {1\over 4}\log n +
\log\left( \Gamma \left({1\over 4}\right) \left({3\over
4}\right)^{1/4}/\sqrt{\pi}\right)    ,&if $g=3$;\cr
-n \log {g(g-2)\sqrt{g}\over (g-1)^3(g+1)}
-{1\over 2} \log { g-3\over g-1}     ,&if $g>3$.\cr}
\cr}
$$
Note that the logarithmic corrections at the critical point are due to the
appearance of a zero mode, so that in the calculation of the integral
we must keep terms higher than quadratic in the action.
These logarithmic corrections are a manifestation of the long-range
order. Note also that in the ferromagnetic region we took into account
only one of the two saddle points, corresponding to a pure thermodynamic
state.
\vskip 0.3cm
Up to now we treated the random graphs as annealed disorder, meaning that
they
were allowed to participate in the dynamics
on an equal footing with the Ising spins. We can
quench them by employing the replica trick
$$  {\overline {\log Z}} =
\lim_{k\to 0} {{\overline {Z^k}}-1\over k}
\ .
                                              \eqno(12)
$$
To this effect
we introduce a real field with argument on the hypercube in $k$ dimensions,
$\phi\big(\{\sigma\}\big) \equiv \phi\left(\sigma^1,\ldots
,\sigma^k\right)$. Each vertex of a Feynman diagram
will now be  labelled by the
values of   $k$  distinct spins, one for each
replica\footnote{$^{(\star)}$}{\sevenrm Note that upper
indices label the replicas. They should not be confused with lower indices
which
label the $\scriptstyle  2n$ vertices of a graph.}.
Arguing as before we can   express the $k$th moment
of the Ising partition function in zero magnetic field
as follows:
$$
{\overline {Z_{{\cal G}_n}^{\;k}}}  \times {\cal N}_n   =
 {1\over 2\pi i} \oint {d\lambda\over \lambda^{2n+1}}
{1\over \sqrt{\det \Delta }}\!
\int
\prod_{\{\sigma \}}{d\phi\big(\{\sigma\}\big) \over \sqrt{2\pi} }\,
\exp (-S)  \ ,                                       \eqno(13)
$$
with
$$
S = {1\over 2}
\sum_{\{\sigma \} , \{ \tau \}}
\phi \big(\{\sigma\}\big) \,
\Delta^{-1}\big(\{\sigma\} ,\{\tau\}\big)   \,
\phi \big(\{\tau\}\big) - {\lambda \over 6} \sum_{ \{ \sigma\} }
\phi \big(\{\sigma\} \big)^3
                                                       \ . \eqno(14)
$$
Here $\sum_{\{\sigma\}}$ stands for a sum over all possible values of the
$k$ spins $\sigma^a$,
and the $2^k\times 2^k$ propagator matrix has entries
corresponding to the Boltzmann weight of $k$ non-interacting replicas
on an  edge,
$\Delta\big(\{\sigma\},\{\tau\}\big) = \exp\left( J \sum_a \sigma^a
\tau^a\right)  $.
More generally we may allow a propagator
$$
\Delta\big(\{\sigma\},\{\tau\}\big) =
\int     dJ \,\rho (J)\,
e^{J \sum_a \sigma^a \tau^a}
                                  \ ,     \eqno (15)
$$
which amounts to choosing uncorrelated couplings on each edge with some
(arbitrary) distribution  $\rho(J)$.
We may also trade the ${\lambda\over 6} \phi^3$ interaction
for a more general monomial
${\lambda^{M-2} \over M!} \phi^M$
so as to obtain graphs with
fixed connectivity equal to $M$.
Extremizing the (rescaled) action yields finally the saddle-point equations
$$
\phi\big(\{\sigma\}\big) =
{1\over (M-1)!} \sum_{\{\tau\}}
\Delta \big(\{\sigma\},\{\tau\}\big) \,
\phi\big(\{\tau\}\big)^{M-1}
                                                \ . \eqno(16)
$$
The calculation
of integer moments of the partition function is thus reduced
in the thermodynamic limit
to a finite algebraic problem.
\vskip 0.3cm
In order to quench the random graphs we
of course still have to continue analytically to values of $k$ near zero.
To do this one must make an ansatz on the precise pattern of
replica-symmetry
breaking. Full symmetry for instance would imply that the field only
depends on
the fraction of replicas pointing up:
$\phi\big(\{\sigma\}\big)=\phi\big(\sigma ^1 +\cdots + \sigma ^k\big)$.
A first stage of hierarchical breaking on the other hand would correspond
to
the ansatz:
$\phi\big(\{\sigma\}\big)=\phi\big(\sigma^1+ \cdots + \sigma^{{k\over m}},
\ldots ,
\sigma^{k-{k\over m}+1}+ \cdots + \sigma^{k }\big)$. Details on the $k\to
0$
continuation as well as on the resulting free energy can be found in refs
[\DeDominicisa, \DeDominicisb]. Here we would only like to point out that
the
order parameter $\phi\big(\{\sigma\}\big)$ can be related to the
more standard magnetization overlaps by the same kind of argument that
lead us to eq.~(10). Indeed the fraction of vertices with a given value
$\{\sigma\}$ for the spins of the $k$ replicas can be easily seen to be
proportional to  $\hat \phi\big(\{\sigma\}\big)^M$. The definition of
the magnetization overlaps on the other hand is
$$
Q_{\phantom i}^{a_{1} \ldots a_{\ell}}
\equiv \lim_{n\to \infty}{1\over 2n}
\sum_i \overline {
\langle \sigma_i^{a_1}\rangle
\cdots
\langle \sigma_i^{a_{\ell}}\rangle}
                                                \ , \eqno(17)
$$
where $\langle A\rangle$ denotes the thermal average, while ${\overline A}$

stands for the  average over the quenched disorder. It follows by
straightforward manipulations that
$$
Q^{a_{1} \ldots a_{\ell}} ={\sum_{\{\sigma \}}
\sigma^{a_1}
\ldots
\sigma^{a_{\ell}}\,
\hat \phi\big(\{\sigma\}\big)^M
\over    \sum_{\{\sigma \}}
\hat \phi\big(\{\sigma\}\big)^M
}
                                                \ . \eqno(18)
$$
\vskip 0.3cm
Our derivation of equations (16) and (18) is the main point
of this letter. Equations (16) had been derived previously
for spin models on the Bethe lattice [\Mott], and were
later argued to hold for random graphs [\DeDominicisa] because
such graphs have a  tree-like local structure.
In [\DeDominicisa, \Mott]
the relation of $\hat \phi\big(\{\sigma\}\big)$ to the overlaps
differs however from eq. (18).
It is conceivable that this difference can be traced to the
effect of finite loops that are ignored in these references.
In any case, besides being exact, our
novel derivation   elucidates the role of the
natural order parameter $\phi\big(\{\sigma\}\big)$ in such models.
As we have shown, it is
the field generating the
diagrammatic expansion, and whose mean-field equations
yield the instanton that governs the behaviour
of this expansion at large orders.
\vskip 0.3cm
The above analysis can be extended easily to several different
contexts. Together with the constraint
$Q^{a}=0\; \forall a$, eqs (16) are for instance the mean-field
equations for the graph bipartitioning problem [\Fu, \Garey].
Fluctuating connectivity can be also
accomodated  if we trade the monomial interaction with a more general
potential
$V(\lambda\phi) / \lambda^2$. The saddle-point equations then read
$$
\phi\big(\{\sigma\}\big) =
 \sum_{\{\tau\}}
\Delta \big(\{\sigma\},\{\tau\}\big) \,
V'\Big(\phi\big(\{\tau\}\big)\Big)
                                     \ ,            \eqno(19)
$$
where $V^\prime$ denotes the derivative of $V$. Note that the
$\lambda$-integration fixes  now the difference of the numbers
of edges and vertices. Other constraints can be   imposed
by extra contour integrations. Eqs (19) with an exponential potential
$V={e^{\alpha (\phi -1)}\over \alpha }$ have been also obtained by De
Dominicis
and Mottishaw [\DeDominicisb] in the case of an ensemble of graphs
where  the connectivity is a random variable with Poissonian distribution
of
average $\alpha $. Finally, as it should be evident, Potts or
continuous spins can be introduced by letting the argument of the field
$\phi$
live on the corresponding space.
\vskip 0.3cm
In the special case of fixed ferromagnetic or
antiferromagnetic coupling  $J$, the mean-field
equations (16) with $M=3$ admit an obvious set of (factorized) solutions
$$
\hat \phi\big(\{\sigma\}\big) = 2^{1-k}
{\hat \phi}_{\sigma_1} \ldots {\hat \phi}_{\sigma_k}
                                                      \ , \eqno(20)
$$
where each factor on the right-hand side stands for (any)
solution of the $k=1$ (annealed)  problem. When the
saddle point (20) dominates,
both the overlaps   and the leading exponential  piece of
${\overline {Z_{{\cal G}_n}^{\;k}}}$
factorize, so that despite the average over  graphs  the replicas are
completely decorrelated\footnote{$^{(\dag)}$}{\sevenrm
Decorrelated   groups  of replicas would correspond more generally to
a product solution  $\scriptstyle  \hat \phi\left(\{\sigma\}\right) =
2^{1-m}
{\hat \phi}_{(k_{1})} \ldots {\hat \phi}_{(k_{m})} $,
where
$\scriptstyle {\hat \phi}_{(k_{\nu })} $
is any solution of the saddle-point equations with $\scriptstyle k_{\nu }$
replicas and $\scriptstyle \sum_{\nu =1}^{m}k_{\nu }=k$. Such solutions
break the symmetry of replicas and are never dominant for integer
$\scriptstyle k$.}. Continuing $k\to 0$ one finds a quenched free energy
equal to
the annealed one, eq. (11),  up to finite-size
corrections\footnote{$^{(\ddag)}$}{\sevenrm
It can be verified more generally under the assumption of replica
symmetry that the factorized solution (20) is indeed dominant in the
$\scriptstyle k\to 0$ limit.}. The corresponding entropy per spin is
$$
\overline{s} = \cases{
{1\over 2}\log{(g+1)^{3}\over 2}
-{3g\over 2(g+1)}\log{g}
,&if $g<3$;\cr
{1\over 2}\log{(g-1)^{3}(g+1)\over g-2}
-{3g(g^2-2g-1)\over 2(g-2)(g-1)(g+1)}\log{g}
,&if $g>3$.\cr}
                                              \eqno(21)
$$
It becomes negative below $g\simeq 0.211$, signaling the existence of
a phase transition in the low-temperature  antiferromagnetic region.
This is also confirmed by an analysis of the moments
${\overline {Z_{{\cal G}_n}^{\;k}}} $ of the partition function.
By solving completely equations~(16) ($M=3$) for $k=2,\ 3$ and $4$ we have
found transition points
$g_{c}^{(2)}\simeq 0.172$, $g_{c}^{(3)}\simeq 0.187$ and $g_{c}^{(4)}\simeq
0.205$, below which the factorizable  saddle point  (20) ceases to
dominate,
so that
$\lim_{n\to \infty}{1\over 2n}\log{\overline {Z_{{\cal G}_n}^{\;k}}}
\not=
\lim_{n\to \infty}{k\over 2n}\log \overline{ Z_{{\cal G}_n}}  $.
This situation is reminiscent of the random-energy model
[\Derrida], except that  the critical temperatures seem to accumulate
to a finite value ($g<1$).  The nature of this low-temperature
phase deserves some further study. Indeed, although the couplings
are purely antiferromagnetic, there is both frustration and disorder since
the random graph has loops of arbitrary size.
\vskip 0.3cm
We conclude with some comments on the interpretation of random graphs
as infinite-genus triangulations. This comes about by considering the
real field $\phi$ as a $N\times N$ hermitean matrix with $N=1$, so that
our ensemble consists of ``fat" graphs ${\cal G}_{n}$ or dual
triangulations
${\cal G}_{n}^*$ [\David] weighted equally for all genera. The average
Euler
characteristic can be computed easily by taking a derivative with respect
to the
size $N$ of the hermitean matrix with the result:
$$
\overline{ \chi} = - n + \log {6n} -
{\partial \log \Gamma (x)\over \partial\log x}\bigg\vert_{x=1}
 \ .                                 \eqno(22)
$$
Note that since for vacuum $\phi ^{3}$ graphs
with $2n$ vertices
$\chi=-n+ \#\, {\rm faces}\, $, an average
graph in this ensemble has a maximal density of handles.
Though rather singular, this 2d surface interpretation allows
a mapping of the Ising model on ${\cal G}_{n}$, onto a model with spins
lying on the vertices of the dual triangular net ${\cal G}_{n}^*$.
This  duality is implemented by a linear transformation of the
fields that diagonalizes the quadratic part of the action.
For $k=1$ for instance the action would take the form
$$
S = {1\over 2} \left({\tilde\phi}_+^2 +
{\tilde\phi}_-^2\right)   - {\tilde \lambda \over 2}
 \left({{\tilde g} \over 3}\, {\tilde\phi}_+^3 -
{\tilde\phi}_+^{\phantom 2} \,{\tilde\phi}_-^2 \right)\ ,
                                          \eqno(23)
$$
with
$$
{\tilde g} = {g+1\over g-1} \ .             \eqno(24)
$$
Since the propagator is now diagonal, we can assign a sign $\pm$
to each edge of the ${\tilde\phi}^3$ graph, or equivalently to the dual
edge $\langle ij\rangle$ on the triangular lattice ${\cal G}_{n}^*$. We
interpret this sign as the value
of $\sigma _i \sigma _j$, where the $\sigma$'s now stand for the
spins residing on the vertices of the triangular lattice.
The product of three signs around a
triangle should be $+$, consistently with the fact that
only two kinds of vertices survive in the action (23).
Furthermore, there is an
extra weight ${\tilde g}$ when all three spins around the
triangle are aligned. As can be verified easily, the duality transformation
(24)
maps the high- and low-temperature ferromagnetic regions of
the Ising models on
${\cal G}_{n}^{\phantom *}$ and
${\cal G}_{n}^*$
to one another. The fact that mean-field theory is exact can be
understood in the dual language as a consequence of the fact that
the number of vertices grows only logarithmically with $n$,
while the connectivity is extensive.
Note finally that the antiferromagnetic region on  ${\cal G}_{n}^*$
corresponds to ${\tilde g}\in [\, 0,1)$, and
is mapped onto the interval
$(-\infty, -1]$. The analysis of the moments and entropy
 shows no signal for a phase transition
in this region.
\vskip 0.6cm
\centerline{\bf Acknowledgements}
\vskip 0.3cm
We have benefited from discussions with C. De Dominicis, M. M\'ezard and
N.~Sourlas. D. Johnston has brought to our attention recently
some interesting numerical simulations of the Ising model on quenched
planar graphs [\John]. The preoccupations in this work are rather
different from ours. This research was supported partially by EEC grant
CHRX-CT93-0340.
\vskip 0.6cm
\centerline{\bf References}
\vskip 0.3cm
\item{[\Bray]}{L. Viana and A.J. Bray, J. Phys. {\bf C18} (1985) 3037.}
\vskip 0.05cm
\item{[\Mez]}{M. M\'ezard and G. Parisi, Europhys. Lett. {\bf 3} (1987)
1067;\hfil\break
I. Kanter and H. Sompolinsky, Phys. Rev. Lett. {\bf 58} (1987) 164.}
\vskip 0.05cm
\item{[\Fu]}{Y. Fu and P.W. Anderson, J. Phys. {\bf A19} (1986)
1605;\hfil\break
J.R. Banavar, D. Sherrington and N. Sourlas, J. Phys. {\bf A20} (1987) L1.}
\vskip 0.05cm
\item{[\DeDominicisa]}{C. De Dominicis and Y.Y. Goldschmidt, J. Phys. {\bf
A22} (1989) L775.}
\vskip 0.05cm
\item{[\DeDominicisb]}{C. De Dominicis and P. Mottishaw, {\sl Sitges Conf.
Proc.}, May 1986, Lecture Notes in Physics {\bf 268}, ed. L. Garrido,
Springer, Berlin and J. Phys.   {\bf A20} (1987) L375.}
\vskip 0.05cm
\item{[\Garey]}{M.R. Garey and D.S. Johnson, {\sl Computers and
Intractability},
Freeman, San Francisco 1979;\hfil\break
C.H. Papadimitriou and K. Steglitz, {\sl Combinatorial Optimization},
Prentice-Hall, Englewood Cliffs, NJ  1982.}
\vskip 0.05cm
\item{[\Par]}{G. Parisi, Phys. Rev. Lett. {\bf 43} (1979) 1754 and
 J. Phys. {\bf A13} (1980) L115, 1101 and 1887;\hfil\break
M. M\'ezard, G. Parisi, N. Sourlas, G. Toulouse and M. Virasoro, Phys. Rev.
Lett.
{\bf 52} (1984) 1156 and J. Physique {\bf 45} (1984) 843; \hfil\break
M. M\'ezard, G. Parisi and M.-A. Virasoro,
 {\sl Spin glass theory and beyond},
 World Scientific, Singapore 1987, and references therein.}
\vskip 0.05cm
\item{[\David]}
{J. Ambj{\o}rn, B. Durhuus and J.
Fr{\"o}hlich, Nucl. Phys. {\bf B257}  [FS14] (1985) 433;
\hfil\break
F. David, Nucl. Phys. {\bf B257} [FS14] (1985) 45 and 543;
\hfil\break
V.A. Kazakov, Phys. Lett. {\bf 150B} (1985) 282;
\hfil\break
V.A. Kazakov, I.K. Kostov and A.A. Migdal,
Phys. Lett. {\bf 157B} (1985) 295.}
\vskip 0.05cm
\item{[\Kaz]} {M. Bershadsky
and A.A. Migdal, Phys. Lett.
{\bf 174B} (1986) 393;\hfil\break
V.A. Kazakov, Phys. Lett.
{\bf 119A} (1986) 140;\hfil\break
D. Boulatov and V.A. Kazakov, Phys. Lett.
{\bf 186B} (1987) 379.}
\vskip 0.05cm
\item{[\Zinn]}{ J.-C. Le Guillou and J. Zinn-Justin
eds, {\sl Large Order Behaviour of Perturbation Theory},
North-Holland, Amsterdam 1989.}
\vskip 0.05cm
\item{[\Mott]}{P. Mottishaw, Europhys. Lett. {\bf 4} (1987) 333.}
\vskip 0.05cm
\item{[\Derrida]}{B. Derrida, Phys. Rev. {\bf B24} (1981) 2613.}
\vskip 0.05cm
\item{[\John]}{C.F. Baillie, K.A. Hawick and D.A. Johnston,
LPTHE-Orsay-94/07
preprint (February 1994).}
\vfil
\bye